\documentclass[12pt]{article}
\usepackage{graphicx}
\usepackage{cite}
\newcommand{\be}{\begin{equation}}
\newcommand{\ee}{\end{equation}}
\newcommand{\beq}{\begin{eqnarray}}
\newcommand{\eeq}{\end{eqnarray}}
\newcommand{\ba}{\begin{array}}
\newcommand{\ea}{\end{array}}
\newcommand{\crs}{cross section}
\newcommand{\crss}{\crs s}
\newcommand{\diff}{differential}
\newcommand{\photo}{photoproduction}
\newcommand{\fsi}{final-state interaction}
\newcommand{\dsdo}{d\sigma/d\Omega}
\newcommand{\gd}{\gamma d}
\newcommand{\gpinn}{\gamma N\to\pi^0 N}
\newcommand{\gpin}{\gamma n\to\pi^0 n}
\newcommand{\gpip}{\gamma p\to\pi^0 p}
\newcommand{\gpid}{\gamma d\to\pi^0 pn}
\newcommand{\getnn}{\gamma N\to\eta N}
\newcommand{\getd}{\gamma d\to\eta pn}
\newcommand{\gnpin}{\gamma N\!\to\!\pi N}
\newcommand{\sgp}{\sigma_{\gamma p}}
\newcommand{\sgd}{\sigma_{\gamma d}}

\newcommand{\rptt}{\rule{0pt}{16pt}}
\newcommand{\ega}{E_{\gamma}}
\newcommand{\qga}{q_{\gamma}}

\newcommand{\kga}{k^{}_{\gamma}}

\newcommand{\half}{\frac{1}{2}}
\newcommand{\lrarr}{\leftrightarrow}
\newcommand{\bfp}{\mbox {\boldmath $p$}}
\newcommand{\bfq}{\mbox {\boldmath $q$}}
\newcommand{\bftau}{\mbox {\boldmath $\tau$}}
\newcommand{\bfsig}{\mbox {\boldmath $\sigma$}}

\newcommand{\bfeta}{\mbox {\boldmath $\eta$}}
\newcommand{\bfgpinn}{\mbox {\boldmath $\gpinn$}}
\newcommand{\bfdeL}{\mbox {\boldmath $\Delta(1232)3/2^+$}}

\date{}

\begin{document}
%\large
\title{\normalsize\bf On the Extraction of Cross Sections for $\pi^0$
	and $\eta$ Photoproduction off Neutrons from Deuteron Data}
\author{ V.E.~Tarasov$^1$, W.J.~Briscoe$^2$, M.~Dieterle$^3$,
	B.~Krusche$^3$,\\
	A.E.~Kudryavtsev$^{1,2}$, M.~Ostrick$^4$, I.I.~Strakovsky$^2$}

%\small
%\date{\today}
\maketitle
\vspace{-7mm}
\centerline{\small $^1$\it {Institute of Theoretical and Experimental
	Physics, Moscow, Russia}}
\centerline{\small $^2$\it {The George Washington University, Washington,
	DC 20052, USA}}
\centerline{\small $^3$\it {Department of Physics, University of Basel,
	Ch-4056, Basel, Switzerland}}
\centerline{\small $^4$\it {Institut f\"ur Kernphysik, Johannes
	Gutenberg-Universit\"at, Mainz, Mainz, Germany}}

\vspace{-4mm}
\begin{abstract}
We discuss the procedure of extracting the \photo\ \crs\ for
neutral pseudoscalar mesons off neutrons from deuteron data. The
main statement is that the \fsi\ (FSI) corrections for the
proton and neutron target are {\it in general} not equal, but
for $\pi^0$ production there are special cases were they have to be
identical and there are large regions in the parameter space of 
incident photon energy and pion polar angle, $\theta^\ast$, 
where they happen to be quite similar.
The corrections for both target nucleons are practically identical
for $\pi^0$ production in the energy range of the $\Delta(1232)3/2^+$
resonance due to the specific isospin structure of this excitation.
Also above the $\Delta$-isobar range large differences between proton 
and neutron correction factors are only predicted for extreme forward 
angles ($\theta^\ast <$~20$^{\circ}$), but the results are similar 
for larger angles. The case of $\eta$ photoproduction is also shortly 
considered. Numerical results for the $\gpip$ and $\gpin$ correction 
factors are discussed. Also the model description for the available 
data on the \diff\ $\gpid$ \crss\ are given.
\end{abstract}

%%111111111111111111111111111111111
\vspace{5mm}
\centerline{\bf 1.~Introduction}
\vspace{2mm}

We analyze the procedure of extracting $\pi^0/\eta$-\photo\
\crss\ off neutrons and off protons from deuteron data. In both cases,
this procedure takes into account the fact that the elementary $\gamma N$
reaction can occur on the proton as well as on the neutron of
the deuteron, and that there are effects from the \fsi\ (FSI).
Thus, even in quasi-free kinematics (close to the kinematics of the
process on the free nucleon), there can be corrections due to this
nuclear effects.
In the case of \photo\ off the proton, one can compare the \crss\ for
free protons to that extracted from the reaction off quasi-free
protons bound in the deuteron. Thus, in the proton case, there is a
possibility to verify the calculation of the corrections.
This verification procedure is impossible for the
reaction off the neutron, since free neutron targets do not exist.
The strategy for the extraction of free-neutron cross-section data
from quasi-free reactions off the deuteron is then to test the FSI
modelling for the proton case and apply the same model to the neutron.
The aim of the present note is to employ a specific approach to the
FSI corrections for the deuteron data to obtain and compare the
correction factors for the elementary \photo\ processes on the
proton and neutron targets in wide regions of initial photon 
energy and of outgoing-pion angles.

%The aim of the present note is to line out and test a specific
%approach of the FSI corrections with deuteron data and free proton
%data.

%%%%%%%%%%%%%%%%%%%%%%%%%%%%%%%%%%%%%%%%%
\begin{figure}%[t]
\begin{center}
\includegraphics[height=3.0cm, keepaspectratio]{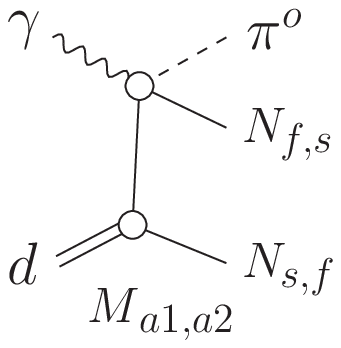} ~~~~~~~
\includegraphics[height=3.0cm, keepaspectratio]{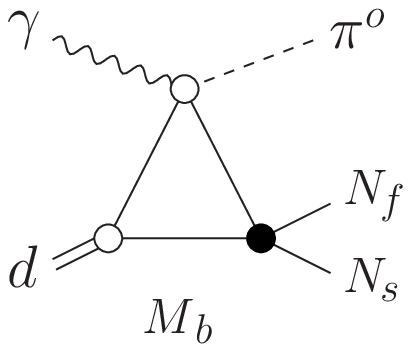} ~~~~~~~
\includegraphics[height=3.0cm, keepaspectratio]{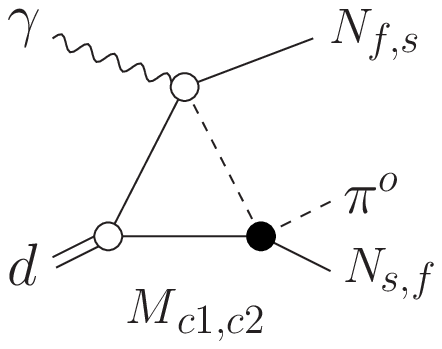} %~~~~~~~~
\end{center}
\vspace{-4mm}
\caption{The impulse-approximation ($M_{a1}$, $M_{a2}$), $NN$-FSI
        ($M_b$) and $\pi N$-FSI ($M_{c1}$, $M_{c2}$) diagrams for
        the reaction $\gpip$. See text for definition of ``fast" ($N_f$)
        and ``slow" ($N_s$) final-state nucleons.}
%  Wavy, solid, dashed and double lines correspond to the photon,
%  nucleons, $\pi^0$ meson and deuteron, respectively.}
        \label{diag}
\end{figure}

%22222222222222222222222222222222222222222222
\vspace{5mm}
\centerline{\bf 2.~Reaction Amplitude}
\vspace{2mm}

Let us write down the $\gpid$ amplitude (the case of $\eta$
production will be discussed later on) as
\be
	M_{\gamma d}=M_{a1}+M_{a2}+M_b+M_{c1}+M_{c2},
\label{1}\ee
where the terms in the r.h.s. are represented in Fig.~\ref{diag}.
The amplitude $M_{\gamma d}$ contains the impulse-approximation
(IA)~\cite{IA} terms $M_{a1}$ and $M_{a2}$ and FSI terms, where
$M_b$ is the $N\!N$-rescattering, while $M_{c1}$ and $M_{c2}$ are
the $\pi N$-rescattering amplitudes.
Hereafter, we consider the kinematics with fast and slow final state
nucleons $N_f(\bfq_f)$ and $N_s(\bfq_s)$, where $\bfq_{f,s}$ are
their momenta in the laboratory system, and $|\bfq_f|\gg|\bfq_s|$.
First, let us discuss the elementary process $\gamma N\to\pi N$.
In the isospin basis, the amplitude has the form~\cite{Berends}
\be
	A(\gamma N\!\to\!\pi_a N)=
	A_v\delta_{a3}+A_{v1}\half [\tau_a,\tau_3]+A_s\tau_{a\,},
\label{2}
\ee
where $\tau_a$ is the Pauli matrix; $A_v$ and $A_{v1}$ are two
isovector amplitudes and $A_s$ is the isoscalar one.
For the proton and neutron targets, we have
\be
	A(\gpip)= A_v+A_s,~~~~ A(\gpin)= A_v-A_s,
\label{3}
\ee
i.e., in the general case, when $A_s\ne 0$, the $\gamma p$ and $\gamma n$
amplitudes are not equal.
Now, we shall consider two different kinematics. Case~1 corresponds to
a fast proton and a slow neutron in the final state, i.e., $N_f\!=\!p$
and $N_s\!=\!n$.
We define case~2 as the result of the isospin replacement $p\lrarr n$
for case~1, i.e., $N_f\!=\!n$ and $N_s\!=\!p$, with identical momenta
and spins of the fast and slow nucleons.  Then, the IA amplitudes $M_{a1}$
and $M_{a2}$ read
\be
\ba{l}
	M_{a1}=[(A^{(1)}_v+A^{(1)}_s)\,\varphi_d(\bfq_s)],~~
	M_{a2}=[(A^{(2)}_v-A^{(2)}_s)\,\varphi_d(\bfq_f)]~~ ({\rm case~1});
\\ \rptt
	M_{a1}=[(A^{(1)}_v-A^{(1)}_s)\,\varphi_d(\bfq_s)],~~
	M_{a2}=[(A^{(2)}_v+A^{(2)}_s)\,\varphi_d(\bfq_f)]~~ ({\rm case~2}).
\ea
\label{4}
\ee
where $\varphi(\bfq)$ is the deuteron wave function (DWF). Hereafter
in this paper, we consider only the isospin structure of the amplitudes
and use the short notation $[\cdots]$, which means that all particle
momenta and spins are properly taken into account. More detailed
expressions with all the variables written out can be found, e.g., in
Refs.~\cite{FSI,Fix,Lev} (see also references therein). The superscripts
``(1)" and ``(2)" of the terms $A^{(1)}_{v,s}$ and $A^{(2)}_{v,s}$ mean
that they are taken at different values of spin and momentum variables
due to the replacement $N_f\lrarr N_s$, i.e.,
$A^{(1)}_{v,s}\equiv\!\!\!\!\!/\,\,A^{(2)}_{v,s}$.  Both amplitudes
$M_{a1}$ and $M_{a2}$ change from case~1 to case~2 because the relative
sign of the term $A_s$ changes.  Since $|\bfq_s|\ll |\bfq_f|$, we have
$\varphi(\bfq_s)\gg\varphi(\bfq_f)$ and $|M_{a1}|\gg |M_{a2}|$. Thus,
the leading diagram is $M_{a1}$, i.e., the reaction $\gpid$ proceeds
mainly through the subprocess $\gpip$ or $\gpin$ in case~1 or 2,
respectively (the detectable recoil nucleon is the participant of
the reaction).

\vspace{2mm}
Now consider the NN-FSI diagram $M_b$ in Fig.~\ref{diag}. First, let
us introduce the $NN$-scattering amplitude in the form
\be
	M(N_1N_2\to N_3N_4)=\half M_0 (\chi^+_3\chi^c_4)(\chi^{c+}_2\chi_1)
       +\half M_1 (\chi^+_3\bftau\chi^c_4)(\chi^{c+}_2\bftau\chi_1).
\label{5}
\ee
Here: $\chi_{1-4}$ are the nucleon isospinors; $\chi^c\equiv\tau_2\chi^*$,
$\bftau=(\tau_1,\tau_2,\tau_3)$, where $\tau_{1-3}$ are the isospin
matrices; $M_0$ and $M_1$ are the isoscalar and isovector $N\!N$
amplitudes. Making use of Eqs.~(\ref{2}) and (\ref{5}), after calculation
with isospin variables one obtains the $N\!N$-FSI amplitude $M_b$ in the
form
\be
\ba{l}
	M_b=\int \,[A_v M_0+A_s M_1]~~~~ ({\rm case~1}),
\\ \rptt
	M_b=\int \,[A_v M_0-A_s M_1]~~~~ ({\rm case~2}),
\ea
\label{6}
\ee
where $\int \,[\cdots]$ denotes the integration over the intermediate
momentum (and sum over intermediate spin states) in the triangle
diagram $M_b$ in Fig.~\ref{diag}. Eqs.~(\ref{6}) show that the
isovector and isoscalar $\gamma N$ amplitudes $A_v$ and $A_s$
couple with I=0 and I=1 $pn$ amplitudes $M_0$ and $M_1$, respectively.
This follows from the simple consideration that for the final $\pi^0(pn)$
system with total I= 0~(1) the $pn$ system should be in an I = 1~(0) state.
From Eq.~(\ref{5}), one can get ``elastic" and ``charge-exchange" $pn$
amplitudes as
\be
	M^{el}_{pn}=M(pn\!\to\! pn)=\half(M_1+M_0),~~~~
	M^{cex}_{pn}=M(np\!\to\! pn)=\half(M_1-M_0),
\label{7}
\ee
and rewrite Eq.~(\ref{6}) in the more illustrative form
\be
\ba{l}
	M_b=\int \,[(A_v\!+\!A_s) M^{el}_{pn}-(A_v\!-\!A_s) M^{cex}_{pn}]~~~~
	({\rm case~1}),
\\ \rptt
	M_b=\int \,[(A_v\!-\!A_s) M^{el}_{pn}-(A_v\!+\!A_s) M^{cex}_{pn}]~~~~
	({\rm case~2}).
\ea
	\label{8}
\ee
The amplitude $M_b$ is represented by two terms. In case~1, the
first (second) term contains the \photo\ amplitude $\gpip$ ($\gpin$)
and subsequent rescattering of the fast proton (neutron) on the slow
neutron (proton). Thus, the first term contains the elastic $pn$ amplitude
$M^{el}_{pn}$ (fast proton $\to$ fast proton), while the second --
charge-exchange one $M^{cex}_{pn}$ involves the fast neutron $\to$ fast proton
scattering. The changes for case~2 are obvious. In both cases, the relative
sign ``$-$" between two terms in Eq.~(\ref{8}) arises from the isospin
antisymmetry of the DWF with respect to the nucleons. In both
Eqs.~(\ref{6}) and (\ref{8}), we again follow only the isospin
structure of the amplitude and show that the only difference between
the results comes from the different relative sign of the term $A_s$ in
cases~1 and 2.
~\vspace{1mm}
Considering the $\pi N$-FSI diagrams $M_{c1}$ and $M_{c2}$ in
Fig.~\ref{diag}, we also obtain for the general case that both these 
amplitudes change from case 1 to case 2 due to the different sign of the 
terms, containing the isoscalar $\gamma N\to\pi N$ amplitude $A_s$.
Hereafter in the present paper, we neglect the $\pi N$-FSI amplitudes,
since their role was found to be small in Refs.~\cite{Fix,Lev,Dar}, and
negligible at energies $E$ above 200~MeV~\cite{Lev}.
\vspace{1mm}
Then the total $\gpid$ amplitude $M_{\gd}$ can be expressed in the form
\be
	M_{\gd} = M_{a1} + \Delta,~~~~ \Delta = M_{a2} + M_b,% +M_{c1}+M_{c2}\;,
\label{12}\ee
where the IA amplitude $M_{a1}$ is the leading-order term (quasi-free
production), and $\Delta$ includes the suppressed IA diagram $M_{a2}$
and the $N\!N$-FSI diagram $M_b$.  In quasi-free kinematics with fast 
and slow final-state nucleons, $\Delta$ is a relatively small correction 
term in comparison to the main one $M_{a1}$. In this kinematics the 
relative momentum between the two final-state nucleons is
%large. Therefore, the contribution from $NN$-FSI is suppressed
%(it is only strong for $s$-wave $NN$-rescattering at small relative
%momenta (Migdal-Watson effect~\cite{Migdal})).
large, and the contribution from $N\!N$-FSI is suppressed, i.e., the
Migdal-Watson effect~\cite{Migdal} from $s$-wave $N\!N$-FSI at small relative
momenta is suppressed. The quasi-free kinematics also implies not very
small momentum transfer from the initial photon to the final pion,
and thus, the FSI effect doesn't essentially cancell the IA amplitudes
due to the orthogonality of the final $pn$ state to the wave function
of the initial deuteron. We shall return to this point below in
Sects.~{\bf 4} and {\bf 5}.

%%3333333333333333333333333333333333333333333333333333333333
\vspace{5mm}
\centerline{\bf 3.~Extraction Procedure for $\bfgpinn$ Reactions}
%            $\gamma N\!\to\!\pi^0 N$ reactions}
\vspace{2mm}

We now discuss the procedure of extracting the $\gamma N$-reaction \crs\
from the deuteron data. First, consider case~1 (fast final-state proton,
slow neutron). In quasi-free kinematics one has the relation~\cite{BL}
(more details are given in the Appendix)
\be
	\frac{d\sigma}{d\Omega}(\gpip)=\frac{1}{n(\bfq_{s\!})}\,
	\frac{d\sigma(\gamma d)}{d\Omega\,d\bfq_s},~~~
	n(\bfq_s)=\frac{\ega'}{\ega}\,\rho(|\bfq_s|).
\label{13}\ee
Here: $d\Omega$ is the solid angle element of the outgoing $\pi^0$ in the
$\pi^0 p_f$ center-of-mass frame with the z-axis directed along the
photon beam; $\ega$ and $\ega'$ are the photon laboratory energies
in the $\gd$ process and the $\gpip$ subprocess, and
$\ega'/\ega\!=1\!+(|\bfq_s|/E_s)\cos\theta_s$, where $E_s$ and
$\cos\theta_s$ are the total energy and polar angle (z-axis along the
photon beam) of the final slow neutron $n_s$ in the laboratory
system. The momentum distribution in the deuteron is given by
$\rho(q) = (2\pi)^{-3}[u^2(q) + w^2(q)]$ ($q = |\bfq|$) where $u(q)$
and $w(q)$ are the $S$- and $D$-wave parts of the DWF. It is normalized
to $\int \!\rho(q)\,d\bfq=1$.
One can improve Eq.~(\ref{13}), taking FSI corrections into account.
With these corrections, we have
\be
	\frac{d\sigma}{d\Omega}(\gpip) = \frac{R_p}{n(\bfq_{s\!})}\,
	\frac{d\sigma(\gd)}{d\Omega\,d\bfq_s}.
\label{14}\ee
The extraction procedure for the $\gpip\,$ reaction implies Eq.~(\ref{14})
to be utilized with the deuteron \crs\ $d\sigma(\gd)/d\Omega\,d\bfq_s$,
taken from the data. Here, the correction factor $R_p$ can be
obtained from the model as
\be
	R_p = \frac{d\sigma^{(p)}(\gd)}{d\Omega\,d\bfq_s}
	\bigg/\frac{d\sigma(\gd)}{d\Omega\,d\bfq_s},
\label{15}\ee
where the \crss\ $d\sigma^{(p)}(\gd)$ is calculated from the main
IA diagram $M_{a1}$, while $d\sigma(\gd)$ is obtained from the full
amplitude $M_{\gd} = M_{a1}+\Delta$ in Eqs.~(\ref{12}) and
should restore the experimental data.
Consider now case~2 (fast neutron, slow proton). Instead of
Eqs.~(\ref{14}) and (\ref{15}), we obtain
\be
	\frac{d\sigma}{d\Omega}(\gpin) = \frac{R_n}{n(\bfq_{s\!})}\,
	\frac{d\sigma(\gd)}{d\Omega\,d\bfq_s},
~~~~~
	R_n = \frac{d\sigma^{(n)}(\gd)}{d\Omega\,d\bfq_s}
	\bigg/\frac{d\sigma(\gd)}{d\Omega\,d\bfq_s},
\label{16}\ee
where $d\Omega$ is the solid angle element of the outgoing $\pi^0$ in
the $\pi^0 n_f$ center-of-mass frame with the z-axis directed along
the photon beam. All other notations are the same as in
Eqs.~(\ref{14}) and (\ref{15}) with the replacements $p_f\!\to\!n_f$
and $n_s\!\to\!p_s$.  Here, the \crs\ $d\sigma^{(n)}(\gd)$ is also
calculated with the main diagram $M_{a1}$, but the fast nucleon is
a neutron.
The correction factors $R_p$ in Eq.~(\ref{15}) (case~1) and $R_n$
in Eqs.~(\ref{16}) (case~2) are calculated through the amplitudes
given by Eqs.~(\ref{4}) and (\ref{6}). Going  from
case~1 to case~2, we change only the relative sign of the isoscalar
\photo\ amplitude $A_s$ in these Eqs. Thus, in the general case
\be
	R_n\ne R_p.
\label{17}\ee
%However, when $A_s = 0$ or $A_v = A_{v1} = 0$ in Eq.~(\ref{2}), we
However, when $A_s = 0$ or $A_v = 0$ in Eq.~(\ref{2}), we get
$R_n = R_p$. In this particular cases, the differential \crss\
of the $\gpip$ and $\gpin$ reactions are equal.
%%%%%%%%%%%%%%%%%%%%%%%%%%%%%%%%%%%%%%%%%%%%%%%%%%%%%%%%%
\begin{figure}%[t]
\begin{center}
\includegraphics[height=5.0cm, angle=90, keepaspectratio]{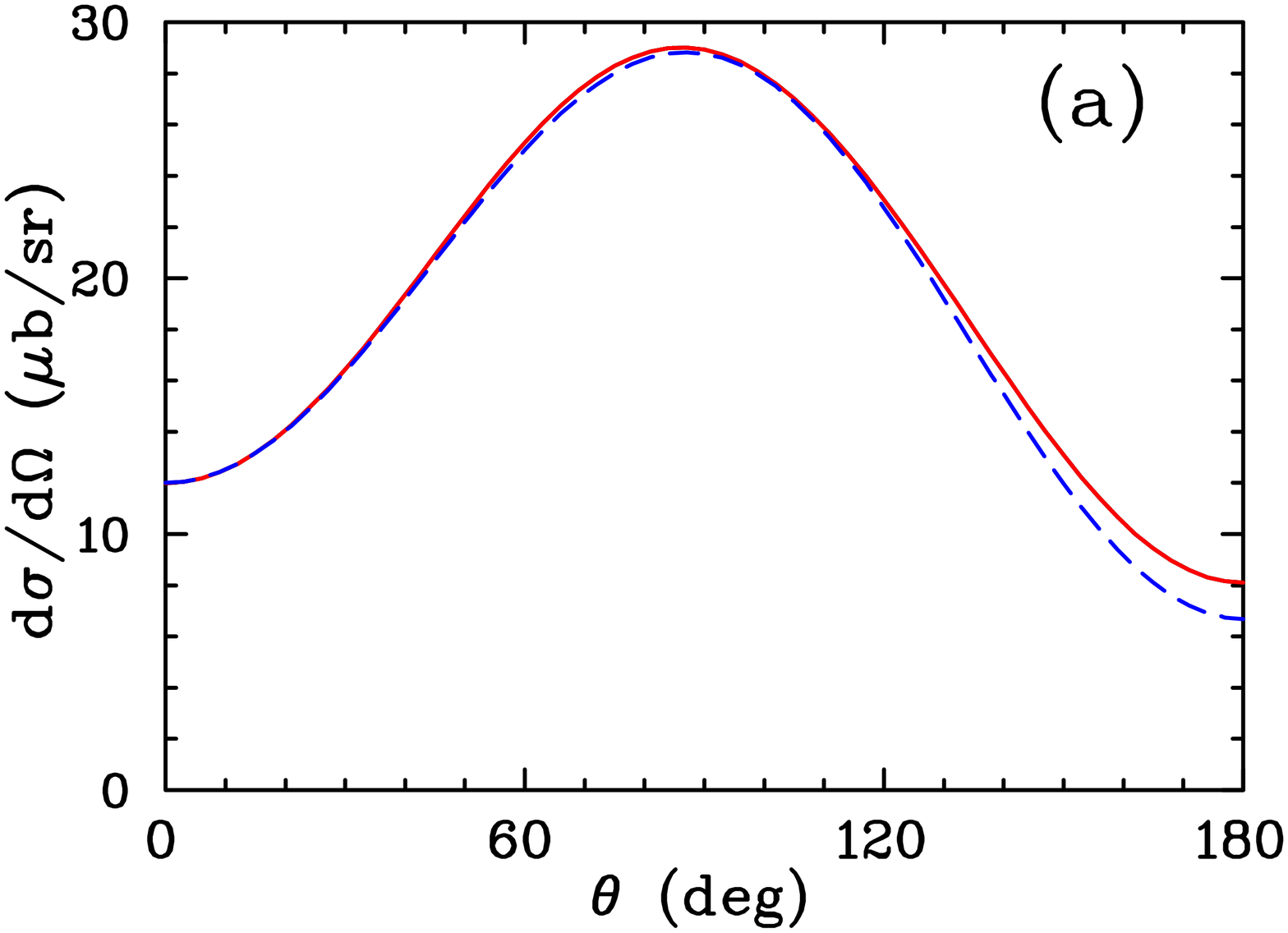}\hfill
\includegraphics[height=5.0cm, angle=90, keepaspectratio]{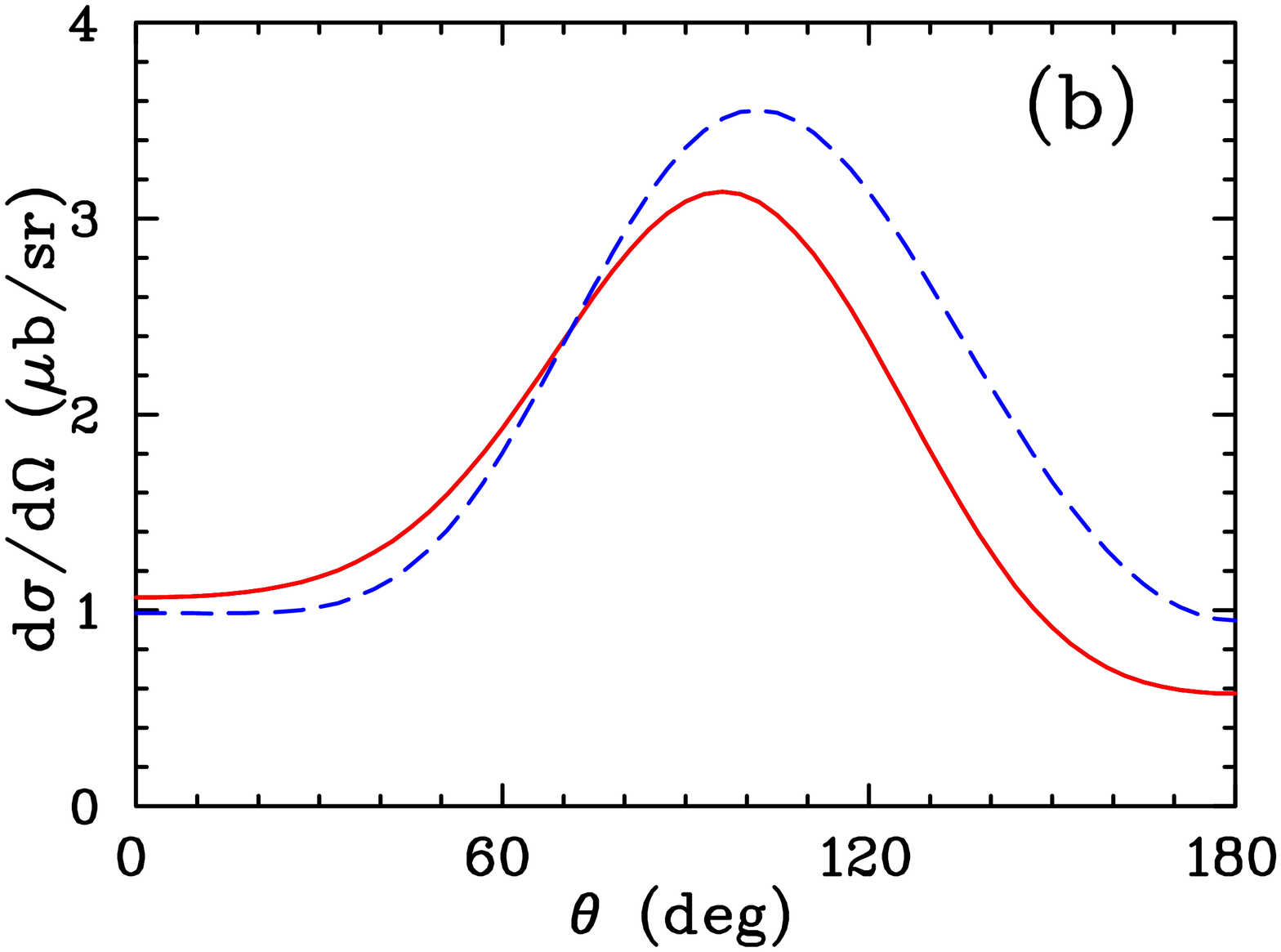}\hfill
\includegraphics[height=5.0cm, angle=90, keepaspectratio]{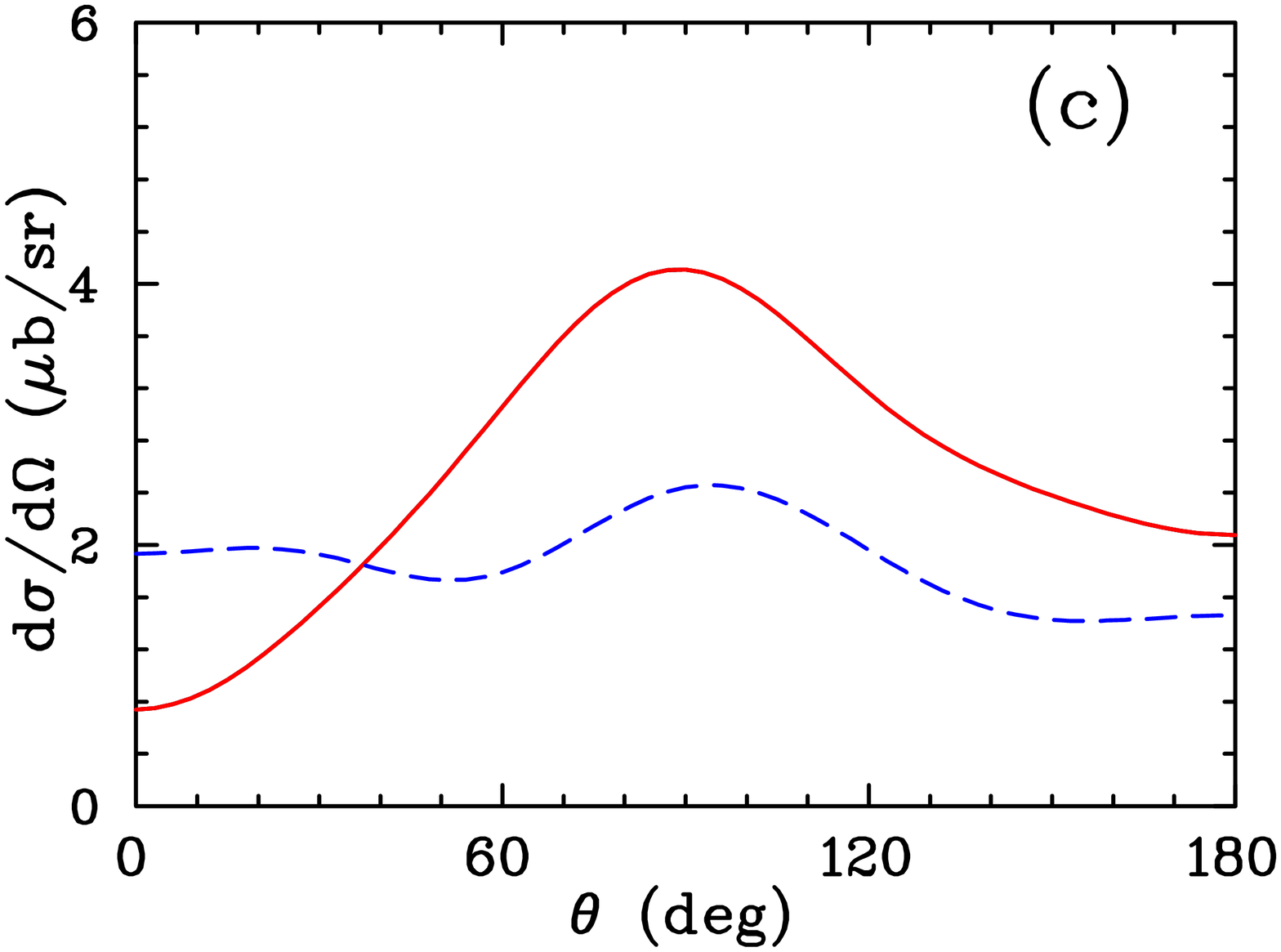}
\end{center}
\vspace{-4mm}
\caption{(Color online) The differential \crss\ of the $\gpip$ (red solid
        curves) and $\gpin$ (blue dashed curves) reactions at several photon
        energies $\ega = 340$ ($a$), 630 ($b$) and 787~MeV ($c$), which
        correspond to $\Delta (1232)3/2^+$, $N(1440)1/2^+$, and
        $N(1535)1/2^-$ regions, respectively.} \label{gpi}
\end{figure}

%3.1%%%3.1%%%3.1%%%3.1%%%3.1%%%3.1%%%3.1%%%3.1%%%3.1%%%3.1%%%
\vspace{3mm}
{\bf 3.1~Comment on the $\bfdeL$ Region}

\vspace{1mm}
Consider the $\pi^0$ \photo\ in the $\Delta (1232)3/2^+$ region.
Supposing that in this region ($\ega\approx 340$~MeV) the $\gpinn$
proceeds through $\Delta(1232)3/2^+$, we obtain $A_s = 0$ due to isospin
conservation. In this case, we get $R_n\! = R_p$, i.e., the
correction factors for the reactions $\gpip$ and $\gpin$ appear to
be the same, and we should also have equal differential \crss\ for
these reactions. To illustrate this, we reconstruct in Fig.~\ref{gpi}
these \crss\ at several energies, obtained with the $\gamma
N\to\pi N$ amplitude taken from the SAID database~\cite{SAID}.
Fig.~\ref{gpi} shows that proton and neutron \crss\ are very close
to each other in the $\Delta(1232)3/2^+$ region ($\ega=340$~MeV). At
higher energies, the contributions from $N(1440)1/2^+$ and
$N(1535)1/2^-$ become important, the isoscalar amplitude
$A_s\ne 0$, and the difference between proton and neutron
differential \crss\ is visible. The main ($M_{a1}$) and
FSI-correction ($\Delta$) terms in the $\gpid$ amplitude
$M_{\gd}$~(\ref{12}) also become different in cases~1
and 2, and we get Ineq.~(\ref{17}).

%%%3.2%%%3.2%%%3.2%%%3.2%%%3.2%%%3.2%%%3.2%%%3.2%%%3.2%%%3.2%%%
\vspace{3mm}
{\bf 3.2~Comment on the $\bfeta$-Photoproduction Case}
\vspace{1mm}

Let us comment on the case of $\eta$ \photo\ on the nucleon ($\getnn$)
and deuteron ($\getd$). The amplitude on the nucleon has isospin
structure $A(\getnn) = A_v\tau_3\! + A_s$, which gives
\be
	A(\gamma p\to\eta p) =A_v+A_s,~~~~ A(\gamma n\to\eta n) =-A_v + A_s,
\label{18}
\ee
The difference with the $\pi^0$ case is that the isovector amplitude
$A_v$ changes sign for $\eta$ production. Thus, the amplitudes $M_{a1}$,
$M_{a2}$, and $M_b$ on the deuteron with the $\pi^0$ replaced by an $\eta$
are given by the same Eqs.~(\ref{4}), (\ref{6}) and (\ref{8}) with
small modifications, i.e., the amplitude $A_v$ (instead of $A_s$)
changes sign when going from case~1 (fast proton) to case~2 (fast
neutron). Thus, we arrive at the same conclusion~(Eq. \ref{17}) in
the general case of $A_v\ne 0$ and $A_s\ne 0$. Addition of $\eta N$-FSI
amplitudes, similar to $M_{c1}$ and $M_{c2}$ in Fig.~\ref{diag}, or
the $\getd$ amplitude with intermediate pion \photo\ followed by the
$\pi N\!\to\!\eta N$ subprocess also lead in general to $R_p\neq R_n$.
However, one should note that in this case FSI effects seem to be much
smaller than for $\pi^0$ photoproduction. The experimental data
\cite{Werthmueller_13,Werthmueller_14} show that $R_p\approx$~1 over a 
wide range of incident photon energies and $\eta$ c.m. polar angles. A 
major difference to $\pi^0$ photoproduction is that this reaction is 
dominated for not too high incident photon energies by the excitation 
of $S_{11}$ resonances via the $E_{0^+}$ spin-flip multipole. This 
means that in quasi-free production off the deuteron, the final state 
nucleons are dominantly in a spin-singlet, while for $\pi^0$ production 
the spin-triplet configuration of the deuteron dominates the final 
state.  This effect (different spin configuration) of initial (deuteron) 
and final $np$-system suppresses the role of FSI in the $\eta$ case in
comparison with the $\pi^0$-production case.
%We plan to discuss this point in more detail later on.
The incoherent $\eta$-photoproduction off the deuteron was studied in
a number of papers (see Ref.~\cite{Fix15} and references therein), and
the FSI effects were found to be important only in the very nearthreshold 
region.  At higher energies, they are quite small~\cite{Fix15}, except 
for special kinematic configurations, where the IA contributions are 
suppressed. 
%The FSI effect were also found to be
%sizeable for some specific polarization variables.
%Some polarization observables were also found there to be rather
%sensitive to $\eta N$ and intermediate-pion ($\pi N\to\eta N$)
%FSI effects.
Theoretical results of Ref.~\cite{Fix15} show visible sensitivity of 
some polarization observables to $\eta N$ and intermediate-pion ($\pi 
N\to\eta N$) FSI effects, but this issue is beyond the scope of the 
present paper.

%44444444444444444444444444444444444444444
\vspace{5mm}
\centerline{\bf 4.~Numerical Results}
\vspace{2mm}

Here, we perform some numerical comparison of the correction factors
$R_p$~(\ref{15}) and $R_n$~(\ref{16}) to illustrate Ineq.~(\ref{17}).
Let us briefly describe the ingredients used in the calculation of
the reaction amplitude $M_{\gamma d}$~(\ref{1}). We use the elementary
$\gamma N\!\to\!\pi N$ amplitudes, generated through the GW pion
\photo\ multipoles~\cite{gN}. For the $NN$-FSI term $M_b$ in
Fig.~\ref{diag}, we include the $s$-wave $pn$-scattering amplitudes
$M_0$ and $M_1$, introduced in Eq.~(\ref{5}). These invariant
amplitudes for $N_1N_2\!\to\!N_3N_4$ read
\be
	M_0 = 8\pi\sqrt{\!s}\,f^{(0)}_{pn}(p)
	(\varphi^+_3\bfsig\varphi^c_4)(\varphi^{c+}_2\bfsig\varphi^{}_1),
~~~
	M_1 = 8\pi\sqrt{\!s}\,f^{(1)}_{pn}(p)
	(\varphi^+_3\varphi^c_4)(\varphi^{c+\!}_2\varphi^{}_1)
\label{19}\ee
($\sqrt{\!s}$ is the effective $N\!N$-system mass).
Here: $\varphi^{}_{1,2,3,4}$ are the nucleon Pauli spinors
($\varphi^+\varphi = 1$); $\varphi^c = \sigma_2\varphi^\ast$. The
amplitude $M_0(M_1)$ with isospin $I\! = \!0(1)$ corresponds to the
total spin $S = 1(0)$. The amplitudes $f^{(0,1)}_{pn}(p)$, where $p
= |\bfp|$ is the relative momentum  in the $N\!N$ system, are
expressed through the scattering lengths $a^{}_{0,1}$ and effective
radii $r^{}_{0,1}$, i.e.,
\be
	f^{(0,1)}_{pn}(p) = \frac{1}{-a^{-1}_{0,1} + \half r^{}_{0,1}
	p^2 - ip},
~~~~~~
	f^{(0,1)\,off}_{pn}(q,p) = \frac{p^2\! + \beta^2_{0,1}}{q^2\!
	+\beta^2_{0,1}}\,f^{(0,1)}_{pn}(p),
\label{20}\ee
where $f^{(0,1)\,off}_{pn}(q,p)$ are the off-shell $N\!N$-amplitudes,
and $q = |\bfq|$ is the relative momentum of the intermediate nucleons.
The off-shell dependence of $f^{(0,1)\,off}_{pn}(q,p)$ is introduced
through the monopole-type formfactor, as given in Eq.~(\ref{20}),
with parameters $\beta_{0,1}$. We take the typical value $\beta_{0,1}
= 1.2$~fm$^{-1}$ (the same as in Refs.~\cite{FSI,Lev}). For $a_{0,1}$
and $r_{0,1}$, we use the known values~\cite{QM}: $a_0 = 5.4$~fm, $r_0
= 1.7$~fm, $a_1 = -24$~fm, and $r_1 = 2.7$~fm.
Hereafter in the present paper, we neglect higher partial waves
($L\!>0$) in the $N\!N$-scattering amplitude. To simplify the calculation
of the $N\!N$-FSI term $M_b$ [Fig.~\ref{diag}], we also take the
amplitude of the $\gpinn$ subprocess out of the loop integral over
the intermediate momentum in this term. This approximation allows
us to calculate analytically the FSI term $M_b$. The more
precise calculations with higher $N\!N$ partial waves included are
planned for the future publication.
%For the $\pi N$-FSI terms $M_{c1,c2}$ in Fig.~\ref{diag}, we include
%the $\pi N$-scattering amplitudes, generated according to the results
%of GW PWA~\cite{piN}.
The DWF was taken from the Bonn potential (full model)~\cite{DWF}.

\begin{figure}%[t]
\begin{center}
\includegraphics[width=14.0cm, keepaspectratio]{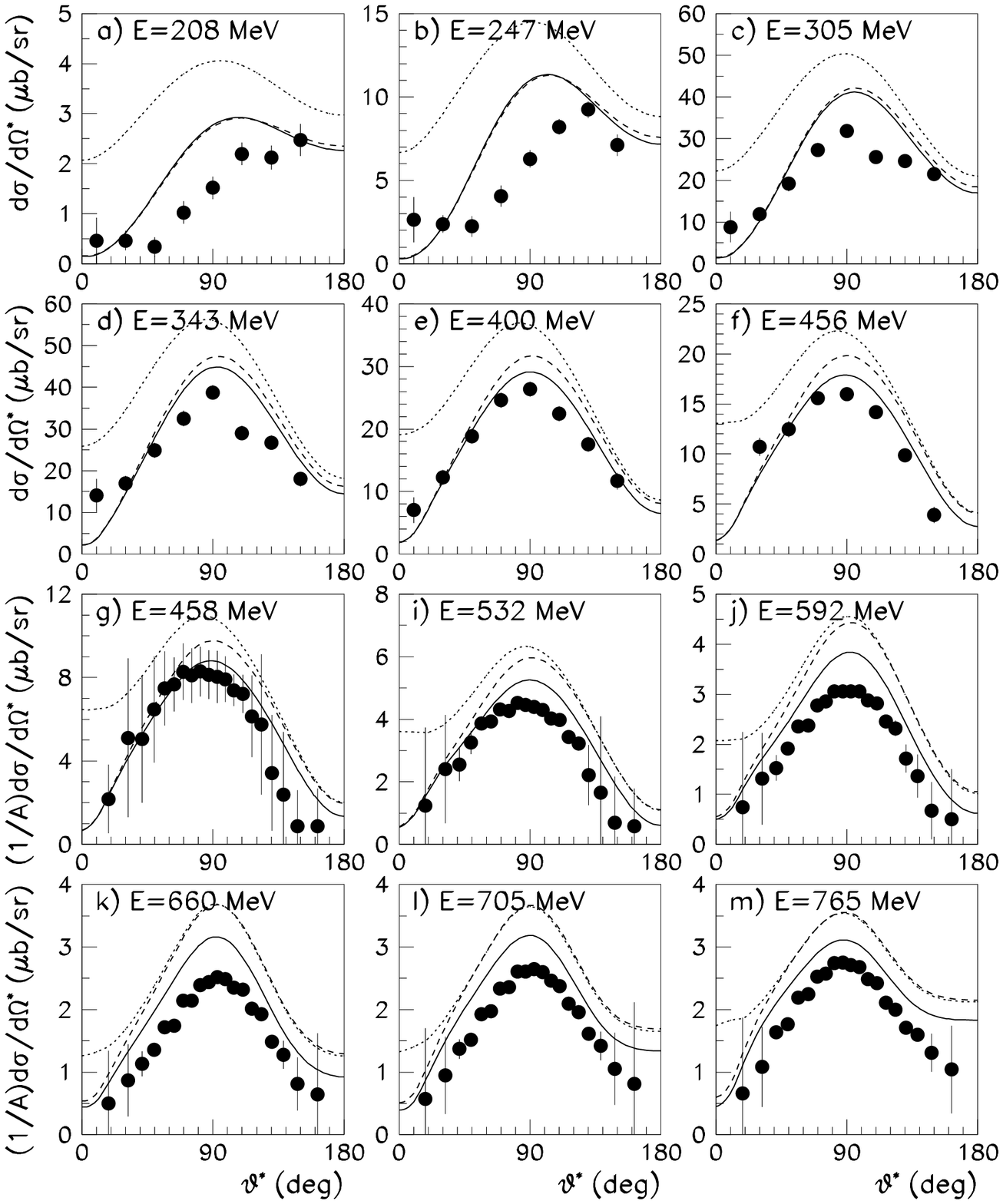}
\end{center}

\vspace{-4mm}
\caption{The differential \crss\ of the $\gpid$ for several values of
 the photon-beam laboratory energy $E$ vs. $\theta^\ast$ (the angle
 $\,\theta^\ast$ is defined in the text).
 The curves show the contributions: dotted -- from the IA amplitudes
 $M_{a1,a2}$ [Fig.~\protect\ref{diag}]; dashed -- from the total
 amplitude $M_{\gd}$~(\ref{12}) with the $N\!N$-FSI.
 Solid curves include also the off-shell correction~(\ref{off})
 for the $\gamma N\!\to\!\pi N$ amplitude.
 The filled circles: in the plots ($a$-$f$) -- the data from 
 Ref.~\protect\cite{Krusche}
 (error bars include statistical uncertainties only); in the plots 
 ($g$-$m$) -- the data
 from Ref.~\cite{Diet} (error bars include statistical and systematic 
 uncertainties in quadratures), multiplied by $1/A$ ($A=2$ for the 
 deuteron).}  \label{s11}

\end{figure}

\vspace{1mm}
The results of the model for differential cross sections $\dsdo^*$
of the reaction $\gpid$ for several photon laboratory energies $E$
are compared in Fig.~\ref{s11} to experimental data from
Ref.~\cite{Krusche} [plots ($a$-$f$)] and to recent data from
Ref.~\cite{Diet} [plots ($g$-$m$)].
Here, $\theta^\ast$ is the polar angle of the outgoing $\pi^0$ in
the c.m. system of the incident photon and a nucleon at rest
%("photon-nucleon" c.m. frame)
with z-axis directed along the photon momentum.
The dotted curves show the contributions from the IA term
$M_a\!=M_{a1}\!+M_{a2}$, while the dashed ones represent the results
obtained with the full amplitude $M_{\gd}\!=M_{a1}\!+M_{a2}\!+M_b$~(\ref{12}).
Comparison of these curves shows an important role of the $N\!N$-FSI,
which essentially decrease the \crss\ at low energies in line with
other papers~\cite{Fix,Lev,Dar}. At higher energies [plots($j$-$m$)],
the FSI effect is very small, except the region of small angles
$\theta^\ast$, where the effect is sizeable.
Our model predictions (dashed curves) sizeably overestimate the data,
but the shape of the \diff\ \crss\ in the main is reproduced.
It is quite possible that the full account of multiple-scattering
diagrams (not only $M_b$ term in Fig.~\ref{diag}) may lead to
destructive interference and decrease the \crss, insignificantly
affecting the shape of the distributions.

Let us try to improve the theoretical description, taking into account 
that the $\gnpin$ amplitudes in the 
diagrams $M_{a1}$, $M_{a2}$ and $M_b$ [Fig.~\ref{diag}] are not on the 
mass shell. Let us multiply the elementary $\gamma N\!\to\!\pi N$ 
amplitude by the off-shell correction factor of the form
\be
        F(\qga,\qga')=\frac{\Lambda^2+\qga^2}{\Lambda^2+\qga'{^{\!2}}},
\label{off}
\ee
where $\qga$ ($\qga'$) is the photon c.m. momentum in the $\gnpin$
reaction, calculated with on-shell (off-shell) initial nucleon at a
given effective $\gamma N$ mass. We add this factor to the IA amplitudes
$M_{a1}$ and $M_{a2}$. For simplicity, we neglect this correction in
the $N\!N$-FSI term, where the momenta of the nucleons in the deuteron
vertex are effectively small in the loop integral, and the off-shell
effect is expected to be small. 
%On the other hand, we expect that
%the IA contributions at high energies [Fig.~\ref{s11}, plots ($j$-$m$),
%dashed curves], where FSI effects are very small, are overestimated
%in the model and should be suppressed. Fig.~\ref{r} shows the
%results from the total amplitude $M_{\gd}$~(\ref{12}).  Here, dotted 
%curves mean the same as in Fig.~\ref{s11}, while dashed ones show the
%off-shell corrected results with $\Lambda=1$~fm$^{-1}$
%in Eq.~(\ref{off}) (we use the value of the same order of magnitude as
%$\beta's$ in Eqs.~(\ref{20})). The role of the off-shell correction is
%negligible at low energies [Fig.~\ref{r}, plots ($a$-$c$)], and visibly
%increases at higher energies. However, varying
%the value of $\Lambda$~(\ref{off}), one can not decrease the predicted
%\crss\ in accordance with the data. Successively applying the cut
%$p^{lab}_n>50$~MeV$/c$ (solid curves), less rigid than in Fig.~\ref{s11},
%we essentially improve the description of the data in the wide energy
%range.  Kinematical cuts for slow particles and off-shell corrections
%for $\gamma N\to\pi N$ amplitudes are rather questionable points but
%leave some room to improve the model predictions.
On the other hand, we expect that
the IA contributions at high energies [Fig,~\ref{s11}, plots ($j$-$m$),
dotted curves], where FSI effects are very small, are overestimated
in the model and should be suppressed.
Solid curves in Fig.~\ref{s11} show the off-shell corrected results
from the full amplitude $M_{\gd}$~(\ref{12}) with $\Lambda=1$~fm$^{-1}$
in Eq.~(\ref{off}) (we use the value of the same order of magnitude as
$\beta's$ in Eqs.~(\ref{20})). The role of the off-shell correction is
negligible at low energies [Fig.~\ref{s11}, plots ($a$-$c$)], but visibly
decreases the \crss\ at higher energies, and the theoretical description
looks better. However, the model still overestimates the data, and varying
the value of $\Lambda$~(\ref{off}), one can not decrease the predicted
\crss\ in accordance with the data. Here we finish the comparison
of the model with the data, leaving further developments, which perhaps
should involve the multiple-scattering diagrams, to the next publications. 
%%%%%%%%%%%%%%%%%%%%%%%%%%%%%%%%%%%%%%%%%
\begin{figure}%[t]
\begin{center}
\includegraphics[width=14.0cm, keepaspectratio]{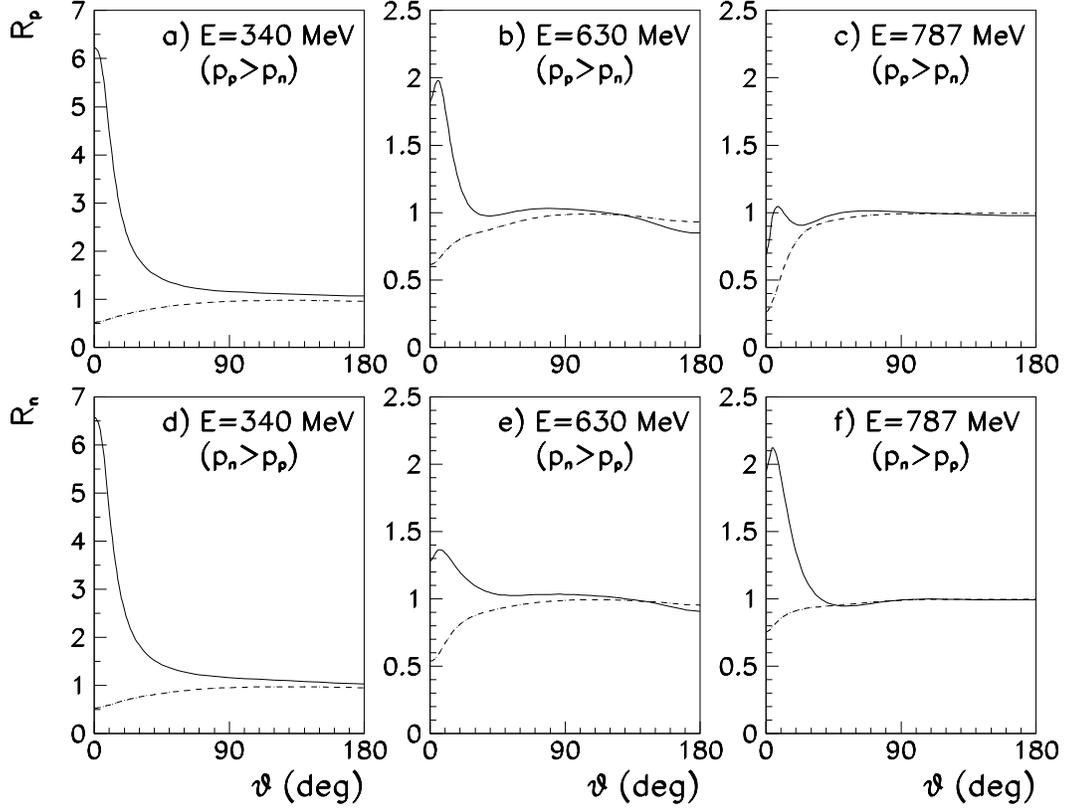}
\end{center}
\vspace{-4mm}
\caption{The correction factors $R_p$ [plots ($a$-$c$)] and $R_n$
        [plots ($d$-$f$)], calculated according to Eqs.~(\protect\ref{21})
        from the reactions $\gamma_d\!\to\!\pi^0 p_f n_s$ and
	  $\gamma d\!\to\!\pi^0 n_f p_s$, respectively. Left, middle 
	and right plots  -- the results for $\ega = 340$, 630, and 
	787~MeV, respectively. The numerators $d\sigma^{(p,n)}/d\Omega$ 
	in Eq.~(\protect\ref{21}) are obtained from the leading IA 
	amplitude $M_{a1}$. Successive addition of the ``suppressed"
	  IA term $M_{a2}$ and $N\!N$-FSI term $M_b$, when calculating 
	the denominators $d\sigma/d\Omega$ in Eqs.~(\protect\ref{21}), 
	leads to dashed and solid curves,  respectively.} \label{r}
\end{figure}

%%%%%%%%%%%%%%%%%%%%%%%%%%%%%%%%%%%%%%%%%%%%%%
\begin{figure}%[t]
\begin{center}
\includegraphics[width=14.0cm, keepaspectratio]{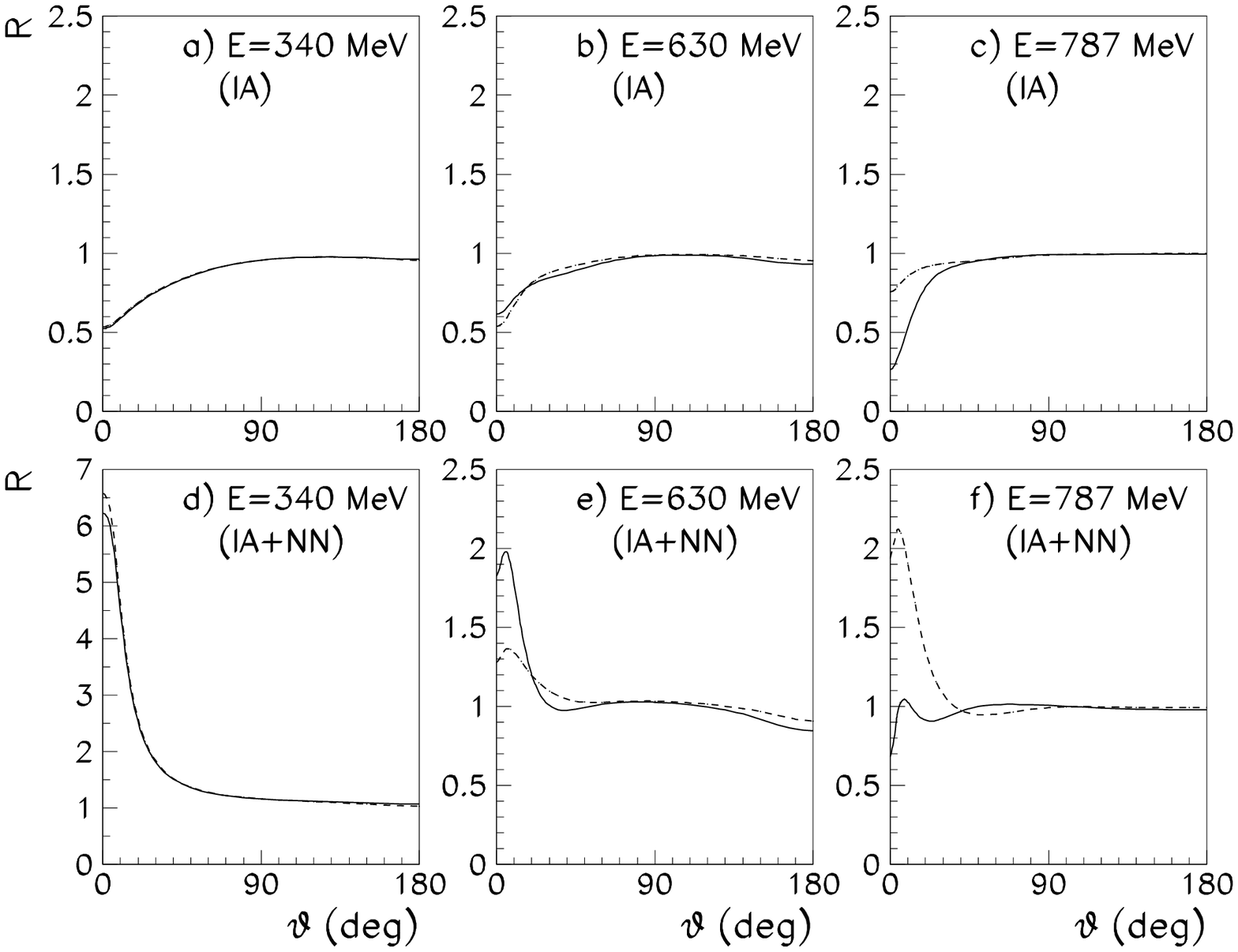}
\end{center}
\vspace{-4mm}
\caption{Comparison of the correction factors $R_p$ (solid curves)
        and $R_n$ (dashed curves) at the same energies $E$ as in
  Fig.~\protect\ref{r}. Upper [($a$-$c$)] and lower [($d$-$f$)] plots
  correspond to the variants shown by dashed and solid curves in
  Fig.~\protect\ref{r}, respectively (see notations therein).}
\label{rr}
\end{figure}

\vspace{2mm}
Now consider the theoretical predictions for the correction factors
$R_p$ and $R_n$, introduced in Sect.~{\b 3}. They are given in
Fig.~\ref{r} at the same photon energies as in Fig.~\ref{gpi}.
For illustrative purposes the results are ``averaged" over the
spectator momentum $\bfq_s$, where $R_{p,n}$ are defined as
\be
	R_p=\frac{d\sigma^{(p)}(\gd)}{d\Omega}\bigg/\frac{d\sigma(\gd)}
	{d\Omega},
~~~~~~
	R_n=\frac{d\sigma^{(n)}(\gd)}{d\Omega}\bigg/\frac{d\sigma(\gd)}
	{d\Omega},
\label{21}\ee
where the differential \crss\ are integrated over the kinematic
region $|\bfq_f|>|\bfq_s|$. Hereafter, we do not apply the off-shell
correction~Eq.~(\ref{off}) to the $\gnpin$ amplitude.
The plots at the top [($a$-$c$)] show the factor $R_p$~(\ref{21})
for case~1 with fast protons and slow neutrons, while the plots
at bottom [($d$-$f$)] show the factor $R_n$ for case~2 with fast
neutrons and slow protons.
The polar angle $\theta^\ast$ of the outgoing pion is defined in the
$\pi^0p$ (case~1) or $\pi^0n$ (case~2) c.m. frame. The types of
curves in Fig.~\ref{r} specify the calculation of the \crss\
$d\sigma(\gd)/d\Omega$, i.e., the denominators in Eqs.~(\ref{21}).
The dashed curves are the results obtained with the full IA amplitude
$M_{a1}\!+\!M_{a2}$  for these \crss, i.e., they include the correction
for the ``suppressed" IA term $M_{a2}$. Addition of the $N\!N$-FSI term
$M_b$ leads to the solid curves. We see that the ``suppressed"
IA term $M_{a2}$ alone already produces a sizeable deviation
$R_{p,n}\ne 1$, which increases to small angles. The $N\!N$-FSI
term $M_b$, when included, considerably affects the results.
Both terms ($M_{a2}$ and $M_b$) essentially affect the results
at small angles $\theta^\ast$ (pions emitted at forward angles), where the
configuration with small relative momenta between the final-state
nucleons dominates, and this $\theta^\ast$ region narrows with the
increasing photon energy. At higher angles, both effects are
negligible, and $R_{p,n}\approx 1$.
At $\ega\!= 340$~MeV [plots ($a,d$)], we obtain a large effect at
$\theta^\ast\sim 0$, where $R_{p,n}\sim 6$. We qualitatively interpret
this results as a self-cancellation effect in the full amplitude
$M_{\gd}\!=M_{a1}\!+M_{a2}\!+M_b$ due to the non-ortogonality
between the initial deuteron state and the final $N\!N$ plane-wave state.
This cancellation should enhance at small momentum transfers
$\Delta$ from the initial photon to the final pion, decreasing the
denominators $d\sigma(\gd)/d\Omega$ in Eqs.~(\ref{21}). The effective
value of $\Delta$ is minimal at $\theta^\ast=0$ and decreases with the
increasing photon energy. We shall also mark this point in the
Conclusion.
%\footnote{$\,$We plan to analyse this cancellation in more details
%         in the future publication.}

The same results are shown in Fig.~\ref{rr} as a comparison of
$R_p$ (solid) and $R_n$ (dashed) factors. The top [($a$-$c$)] and
and bottom [($d$-$f$)] rows show these factors for successive inclusion
of the ``suppressed" IA term $M_{a2}$ and $NN$-FSI term $M_b$ for the
denominators $d\sigma(\gd)/d\Omega$ in Eqs.~(\ref{21}).
At $\ega\!= 340$~MeV, i.e., in the $\Delta (1232)3/2^+$ region, we have
$R_p\!=\!R_n$ to a good accuracy, since the role of the isoscalar
$\gamma N\!\to\!\pi N$ amplitude $A_s$ is negligible.
At higher energies (630 and 787~MeV), we observe that $R_p\!\ne\!R_n$,
because both isovector and isoscalar amplitudes ($A_v$ and $A_s$)
contribute. This difference is considerable at small angles $\theta^\ast$
where the role of the ``suppressed" IA term $M_{a2}$ and $N\!N$-FSI
is enhanced.
Thus, the results in Fig.~\ref{rr} illustrate our analytical
considerations discussed above, which lead to the general
in Ineq.~(\ref{17}).
%(pions emitted at forward angles), where the configuration with small
%relative momenta between the final state nucleons dominates, so that

%555555555555555555555555555555555555555555555
\vspace{5mm}
\centerline{\bf 5.~Conclusion}
\vspace{2mm}

We see that the values $R_{p,n}$ strongly differ from 1 at small
angles $\theta^\ast$. One obtains the large values $R_{p,n}$ when
including the $N\!N$-FSI diagram, which decreases the \diff\ \crss\
at small angles. It means that at small momentum transfers the FSI term
essentially cancels the IA terms due to orthogonality of the final
$pn$-state to the deuteron wave function. This cancellation suppresses
the denominators in Eqs.~(\ref{15}) and (\ref{16}), i.e., the \crss\ on
the deuteron, and leads to the growth of $R_{p,n}$. We plan to return to
this point with more details in the next papers. This effect, not
connected with the Migdal-Watson effect~\cite{Migdal}, was
considered earlier in Ref.~\cite{Ksenz} (see also Ref.~\cite{Fix}).
%\footnote{$\,$We plan to analyse this cancellation in more details
%         in the future publication.}
%
Concerning the physical meaning of the results one should remember
that the photon has no definite isospin. For any given process, it may
be represented as a superposition of I = 0 and I = 1 states, i.e.,
$|\gamma\rangle = c_0 |0\rangle + c_1 |1\rangle$, where the
coefficients $c_{0,1}$ depend on the reaction amplitude. Let us
apply the charge symmetry (CS) operator~\cite{Miller}, which is
the rotation by $180^\circ$ about the y-axis in isospin space with
the z-axis related to the charge, to the process $\gamma d\!\to\!\pi^0
p_f n_s$.  We denote fast and slow protons (neutrons) as $p_{f,s}$
($n_{f,s}$). The CS rotation of the initial and final particles
gives
\be
\ba{l}
	|\gamma\rangle\to|\gamma'\rangle = c_0 |0\rangle\! - \!c_1
	|1\rangle\ne
	|\gamma\rangle,~~~~ |d\rangle\to|d\rangle,~~~~
	|\pi^0\rangle\to - |\pi^0\rangle,~~~
\\ \rptt
	|p\rangle\to |n\rangle,~~~ |n\rangle\to - |p\rangle,~~~
	|pn\rangle_I\to\pm\,|pn\rangle_I~~
%|pn\rangle_{I=1}\to -|pn\rangle_{I=1}.
	(+,-~{\rm for}~I\! = 0,1).
\ea
\label{22}\ee
Then for the amplitudes of interest, we obtain
\be
	M(\gamma d\!\to\!\pi^0 p_f n_s) = M(\gamma' d\!\to\!\pi^0
	n_f p_s) \ne M(\gamma d\!\to\!\pi^0 n_f p_s).
\label{23}\ee
Here the first equality is the result of the CS rotation. The second
inequality means that the corresponding amplitudes are different
because the initial photons are different, i.e.,
$|\gamma\rangle\ne |\gamma'\rangle$. As a result, we obtain
$M(\gamma d\!\to\!\pi^0 p_f n_s)\ne M(\gamma d\!\to\!\pi^0 n_f p_s)$,
and that is the general reason for the Eq.~(\ref{17}).
%\vspace{1mm}
In the $\Delta (1232)3/2^+$ range, where the isoscalar $\gamma N\!\to\!\pi
N$ amplitude $A_s$ is negligibly small, the proton and
neutron correction factors are equal, $R_p = R_n$, to a good
accuracy. This means that in this region one can extract the $\gamma
n\!\to\!\pi^0 n$ differential \crs\ from the $\gamma d\!\to\!\pi^0
n_f p_s$ data by making use of the proton factor $R_p$. This is
convenient, since $R_p$ can be directly obtained from Eq.~(\ref{14}),
making use of the data on the free proton ($\gpip$) and on the deuteron
($\gd\!\to\!\pi^0 p_fn_s$). In this way one can also verify the
model predictions for $R_p$ obtained from Eq.~(\ref{15}).
In the region above the $\Delta(1232)3/2^+$, where both isovector and
isoscalar $\gamma N\!\to\!\pi N$ amplitudes are important, we
have Ineq.~(\ref{17}). The main $R_p/R_n$ differences in our
results are observed at small angles $\theta^\ast$, where contributions
of the ``suppressed" IA amplitude $M_{a2}$ or $N\!N$-FSI terms are
important. However, there is a wide range of large angles $\theta^\ast$,
where we have approximately $R_p\! = \!R_n$.
%\vspace{1mm}
To the end of this paper it is worth saying that the results of
calculations, presented in Figs.~\ref{s11}$-$\ref{rr}, were obtained
not with the full program, but include a number of approximations
with only $s$-wave $N\!N$-rescatterings taken into account and
simplified version of the loop integral in the FSI term $M_b$ as was
mentioned in Sect.~{\bf 4}. We leave more precise full calculations
for the future publication.

%%%%%%%%%%%%%%%%%%%%%%%%%%%%%%%%%%%%%%%%%%%%%%%%%%%%%%%%%%5
\vspace{3mm}
\centerline{\bf Acknowledgements}
\vspace{2mm}

The authors are thankful to V.~V.~Kulikov for useful discussion.
This material is based upon work supported by the U.S. Department
of Energy, Office of Science, Office of Nuclear Physics, under
Award Number DE-FG02-99-ER41110 and the DFG under Grant No. SFB
1044. M.~D. and B.~K. acknowledge support from Schweizerischer 
Nationalfonds.
A.~E.~K. thanks Grant No. NS.3172.2012.2 for partial support.
V.~E.~T. and A.~E.~K. thank the Institute for Kernphysik at Mainz
where part of this work was performed for hospitality and support.

%%%%%%%%%%%%%%%%%%%%%%%%%%%%%%%%%%%%%%%%%%%%%%%%%%%%%%%%%
%\vspace{5mm}
\newpage
\centerline{\bf Appendix:}
\vspace*{5mm}
\centerline{\bf On the Relation between the Cross Sections on the
	Nucleon and Deuteron}

\vspace{2mm}

In quasi-free kinematics, the reaction $\gd\!\to\!\pi^0 p_f n_s$
goes predominantly via the diagram $M_{a1}$ in Fig.~\ref{diag}.
For definiteness, we consider case 1 with a fast proton and a slow
neutron. In this approximation, the \diff\ \crs\ on the deuteron is
related to that on the proton ($\gpip$) through Eq.~(\ref{13}),
which is well-known~\cite{BL}. Here, we give some explanations,
which may be useful when analysing experimental data. Hereafter,
we use the notations
$$
	\kga = (\ega,\bfq_{\gamma}),~~ k^{}_d=(m_d,\vec 0),~~
	k_{\pi} = (E_{\pi},\bfq_{\pi}),~~ k_f=(E_f,\bfq_f),~~
	k_s=(E_s,\bfq_s) \eqno{(\rm A.1)}
$$
for the 4-momenta of the photon, deuteron, pion, fast proton and slow
neutron (spectator), respectively. The total energies and 3-momenta in
the laboratory frame (deuteron rest frame with z-axis along the photon
beam) are given in brackets, and $q_{\gamma}\!=\!|\bfq_{\gamma}|\!
= \!\ega$, $\,q_{\pi}\!=\!|\bfq_{\pi}|$ and $\,q_{f,s}\! =
\!|\bfq_{f,s}|$; $\,m_d$, $m_{\pi}$, $m_p$, $m_n$ are the deuteron,
pion, proton, and neutron masses.
In the unpolarized case one can rewrite Eq.~(\ref{13}) with the full set
of variables as
$$
	\frac{d\sgp}{d\Omega}(W,z;m'^{\,2}_p)\!
	= \frac{1}{n(\bfq_s)}\,\frac{d\sgd}{d\Omega\,d\bfq_s}
	(\ega,z,\varphi,q_s,z_s),
~~~
	n(\bfq_s) = \frac{\ega{\!\!}'}{\ega}\,\rho(q_s).
%n(\bfq_s)=\frac{2 J_{\gamma p}}{J_{\gamma d}}\,\rho(q_s).
%n(\bfp_2)=(1+\beta\cos\theta_2)\rho(p_sa),~~~\beta=\frac{p_s}{E_2}.
	\eqno{(\rm A.2)}
$$
In Eq.~(A.2) is $d\Omega = dz d\varphi$ ($z\! = \cos\theta$) the solid
angle element of outgoing the $\pi^0$ in the $\pi^0 p$ center-of-mass 
frame with
the z-axis along the photon beam, where $\theta$ and $\varphi$ are the
corresponding polar and azimuthal angles. The laboratory polar angle of
the neutron spectator is $\theta_s$ with $z_s\! = \!\cos\theta_s$. $W$ is
the effective $\pi^0 p$ mass related to $\bfq_s$ by
$$
	W^2\! = (\kga+p^{}_d-p_s)^2 =
	(\ega + m_d-E_s)^2\! - \ega^2\! - q^2_s\!
	+ 2\ega q_s z_s~~~ (E_s\! = \!\sqrt{m^2_n\! + q^2_s})
	\eqno{(\rm A.3)}
$$
The momentum-distribution function $\rho(q)$ and photon laboratory
energy $\ega{\!\!}'$ in the subprocess $\gpip$ are defined in connection
with Eq.~(\ref{13}). One can also write $\ega{\!\!}' = (W^2\! - m^2_p)/2m_p$.

For a given $\ega$, the \crs\ $d\sgd/d\Omega\,d\bfq_s$ with unpolarized
particles depends in general case on four variables, chosen in Eq.~(A.2)
as $z$, $\varphi$, $q_s$ and $z_s$. The \crs\ $d\sgp/d\Omega$, being a
function of $W$ and $z$, may also depend on the virtuality $m'^{\,2}_p$ of
the ``target" proton in the subrocess $\gpip$, where $m'^{\,2}_p\! = \!(p_d\!
-\!p_s)^2\! = \!(m_d\!-\!E_s)^2\! - \!q^2_s\ne m^2_p$. Thus, in view
of Eq.~(A.3) the \crs\ $d\sgp/d\Omega$ in Eq.~(A.2) depends only on
three variables, i.e., $q_z$, $z_s$ and $z$. Thus, the \crs\
$d\sgd/d\Omega\,d\bfq_s$ on the deuteron should not depend on $\varphi$
in the model based on the leading IA diagram $M_{a1}$ in
Fig.~\ref{diag}.

For completeness, let us express the cosine $z$ in Eq.~(A.2) through
the particle momenta from Eq.~(A.1). It can be obtained through the
relations
$$
	(\kga k^{}_{\pi}) = \ega^c (E^c_{\pi}\! - \!q^c_{\pi}z),~~
	\ega^c\! = \frac{W^2\!-\!m'^{\,2}_p}{2W},~~
	E^c_{\pi}\! = \!\frac{W^2\! + \!m^2_{\pi}\! - \!m^2_p}{2W},~~
	q^c_{\pi}\! = \!\sqrt{(E^c_{\pi})^2\! - \!m^2_{\pi}},
	\eqno{(\rm A.4)}
$$
where $\ega^c$ and $E^c_{\pi}$ are the total energies of the photon and
pion in the $\pi^0 p$ center-of-mass frame.

%\vspace{2mm}%\item[]
%$J_{\gamma d}$, $J_{\gamma p}$ are the flux factors for the reactions
%$\gpid$ and $\gpip$, respectively, and can be written as
%$$J_{\gamma d}\!=4\ega m_d,~~~~ J_{\gamma p}=4\ega{\!\!}'\,m_p,
%\eqno{(\rm A4)}$$ Then
%$$\frac{2 J_{\gamma p}}{J_{\gamma d}}=\frac{\ega{\!\!}'}{\ega},
%~~~~~\ega{\!\!}'=(W^2\!-m^2_p)/2m_p.\eqno{(\rm A5)}$$

\vspace{2mm}
Let us rewrite Eq.~(A.2) in a form more convenient for applications.
Making the substitutions $d\Omega=dz d\varphi$ and $d\bfq_s\!=q^2_s dq_s
dz_s d\varphi_s$ and averaging Eq.~(A.2) over $\varphi$ and
$\varphi_s$, we obtain
$$
	\frac{d\sgp}{dz}(W,z;m'^{\,2}_p)\! = \frac{1}{2\pi n(\bfq_s)}\,
	\frac{d^{\,3}\sgd}{q^2_s dq_{s\,}dz dz_s}
	(\ega,z,q_z,z_s).
%n(\bfq_s)=\frac{\ega{\!\!}'}{\ega}\,\rho(q_s),
	\eqno{(\rm A.5)}
$$

In the real data analysis, one may divide the phase space of the
reaction $\gpid$ into small cubes, bounded by the values
$$
	z\pm\half\,\Delta_z,~~~~ q_s\pm\half\,\Delta_{q_s},~~~~
	z_s\pm\half\,\Delta_{z_s}.
	\eqno{(\rm A.6)}
$$

Let $\Delta\sigma(\ega,z,q_s,z_s)$ be the $\gpid$ \crs\ in such a
cube~(A.6). Then, we get
$$
	\frac{d\sgp}{dz}(W,z;m'^{\,2}_p)\! = \frac{1}{2\pi n(\bfq_s)}\,
	\frac{\Delta\sigma(\ega,z,q_s,z_s)}{\Delta_z\Delta_{z_s}
	\Delta_{q_s\,}q^2_s},
%(\ega,z,q_z,z_s),
	~~~~ n(\bfq_s)\!=\frac{\ega{\!\!}'}{\ega}\,\rho(q_s),
%~~m'^{\,2}_p\!=(m_d-E_s)^2\!-q^2_s,
	\eqno{(\rm A.7)}
$$
where
$W^2\!=W^2(\ega,q_s,z_s) = \,$Eq.~(A.3),~$\ega{\!\!}' = (W^2\! -
m^2_p)/2m_p$, and $m'^{\,2}_p\! = (m_d-\!E_s)^2\!-q^2_s$. At small
$q_s$ ($\ll m_{p,n}$), one may neglect the $q^2_s$ terms in $m'^{\,2}_p$
and $W^2$~(A.3). Then $m'_p = m_p$, but $W$ depends on $q_s$ and
$z_s$ through the linear term $2\ega q_s z_s$ in Eq.~(A.3), and
$\ega{\!\!}' = \ega[1 + (q_s/E_s)z_s]$.

%%%%%%%%%%%%%%%%%%%%%%%%%%%%%%%%%%%%%%%%%%%%%%%%%%%%%%%%%%%%%
%\newpage

\end{document}